\documentclass[fleqn,twoside]{article}
\usepackage[headings]{espcrc2}

\usepackage{graphicx}
\usepackage[figuresright]{rotating}

\newtheorem{theorem}{Theorem}
\newtheorem{lemma}[theorem]{Lemma}

\def\qed{\hbox{\rlap{$\sqcap$}$\sqcup$}}

\title{Optimal Per-Edge Processing Times in the Semi-Streaming Model\thanks{Supported by the DFG Research Center {\sc Matheon} ``Mathematics for key technologies" in Berlin}}

\author{Mariano Zelke\address{Humboldt-Universit\"at zu Berlin, Institut f\"ur Informatik, 10099 Berlin, Germany}%
        \thanks{Email address: zelke@informatik.hu-berlin.de}}

\begin{document}

\begin{abstract}
We present semi-streaming algorithms for basic graph problems that have optimal per-edge processing times and therefore surpass all previous semi-streaming algorithms for these tasks. The semi-streaming model, which is appropriate when dealing with massive graphs, forbids random access to the input and restricts the memory to ${\cal O}(n\cdot\mbox{polylog}\,n)$ bits. 

Particularly, the formerly best per-edge processing times for finding the connected components and a bipartition are ${\cal O}(\alpha(n))$, for determining $k$-vertex and $k$-edge connectivity ${\cal O}(k^2n)$ and ${\cal O}(n\cdot\log n)$ respectively for any constant $k$ and for computing a minimum spanning forest ${\cal O}(\log n)$. All these time bounds we reduce to ${\cal O}(1)$.

Every presented algorithm determines a solution asymptotically as fast as the best corresponding algorithm up to date in the classical RAM model, which therefore cannot convert the advantage of unlimited memory and random access into superior computing times for these problems.
\vspace{1pc}
\\\emph{Keywords:} graph algorithms, streaming algorithms, per-edge processing time
\end{abstract}

\maketitle

\section{Introduction}
When facing computational tasks on massive graphs the postulate of the classical RAM model, that is, storing the whole input in memory allowing random access to it, is no longer adequate. In fact, information building up the graph may arrive at no specified order and the attempt of completely storing it exceeds common main memories. Regarding this Muthukrishnan\cite{Muthukrishnan} 2003 proposed the \emph{semi-streaming model} as a more restrictive model of computation. According to this the edges of the input graph $G$ appear at arbitrary order and the memory is limited to ${\cal O}(n\cdot\mbox{polylog}\,n)$ bits, where $n$ is the number of vertices in $G$. An important parameter of a semi-streaming algorithm is described by the per-edge processing time $T$, i.e., the time the algorithm needs to handle each single edge. This time determines at which frequency the edges may arrive. The second parameter of the semi-streaming model denotes the number $P$ of passes the algorithm takes over the input stream. All considered algorithms in this paper use only one pass.

Despite the heavy restrictions in the semi-streaming model there are algorithms known solving basic graph problems. In \cite{FeigenbaumKannanMcGregorSuriZhang04} semi-streaming algorithms are given for computing the connected components and a bipartition of a graph as well as a minimum spanning tree of a weighted graph. There are approaches to determine the $k$-edge connectivity  \cite{FeigenbaumKannanMcGregorSuriZhang05} and the $k$-vertex connectivity \cite{FeigenbaumKannanMcGregorSuriZhang05},\cite{Zelke} of a graph for any constant $k$.

In this paper we present semi-streaming algorithms for computing the connected components and a bipartition of a graph, to calculate the $k$-vertex and $k$-edge connectivity for any constant $k$ and to find a minimum spanning forest MSF. All these algorithms have constant and therefore optimal per-edge processing times. 

Section \ref{PrelsAndDefs} gives the usual definitions, in Section \ref{PerEdgeTimeDiscussion} we discuss our definition of the per-edge processing time which is a slight refinement of previous definitions. We develop our semi-streaming algorithms in Section \ref{MainSection}. In Section \ref{Discussion} we debate on  how the obtained algorithms  compete with the corresponding algorithms in the RAM model. A final conclusion is found in Section \ref{Conclusion}.

\section{Preliminaries and Definitions}\label{PrelsAndDefs}
By $G$ we denote a graph $G(V,E)$ with vertex set $V$ and edge set $E$. We call $n=|V|$ and $m=|E|$ the number of vertices and edges respectively. Every graph considered in this paper is undirected and contains no loops but might have multiple edges. For computing an MSF we consider $G$ to be a weighted graph, that is, with a nonnegative weight associated with each edge. Regarding the memory constraints of the semi-streaming model we assume every weight to be storable in ${\cal O}(\mbox{polylog}\,n)$ bits.

We define $\alpha(m,n)$ to be a natural inverse of Ackermann's function $A(\cdot,\cdot)$ as defined in \cite{Tarjan}: $\alpha(m,n):=\min\{i\ge1\;|\;A(i,\lfloor m/n\rfloor)>\log n\}$. We abbreviate $\alpha(n)$ to denote $\alpha(n,n)$.

\ \\
\textbf{Bipartition.} A graph $G$ is called \emph{bipartite} if the vertices can be split in two parts, a \emph{bipartition}, such that no edge runs between two vertices in the same parts. The problem of finding a bipartition is to find two such parts or stating that there is no bipartition since the graph is not bipartite.

\ \\
\textbf{Connectivity.} We name two vertices \emph{connected} if there is a path between them. A graph $G$ is connected if any pair of vertices in $G$ is connected, a \emph{connected component} of $G$ is an induced subgraph $C$ of $G$ such that $C$ is connected and maximal. A \emph{spanning forest} of $G$ is a subgraph of $G$ without any cycles having the same connected components as $G$.
Given a positive integer $k$, a graph $G$ is said to be \emph{$k$-vertex connected} (\emph{$k$-edge connected}) if the removal of any $k-1$ vertices (edges) leaves the graph connected.
A subset $S$ of the vertices (edges) of $G$ we call an \emph{$l$-separator} (\emph{$l$-cut}) if $l=|S|$ and the graph obtained by removing $S$ from $G$ has more connected components than $G$. The \emph{local vertex-connectivity} $\kappa(x,y;G)$ (\emph{local edge-connectivity} $\lambda(x,y;G)$) denotes the number of vertex-disjoint (edge-disjoint) paths between $x$ and $y$ in $G$. By a classical result of Menger (see e.g. \cite{Bollobas}) the local vertex- (edge-) connectivity between $x$ and $y$ equals the minimum number of vertices (edges) that must be removed to obtain $x$ and $y$ in different connected components.

\ \\
\textbf{MSF/MST.} For an edge-weighted graph $G$ the \emph{minimum spanning forest} MSF is a subgraph $G'$ of $G$ with minimum total cost consisting of the same connected components as $G$. If $G$ is connected we name $G'$, which is then connected as well, the \emph{minimum spanning tree} MST of $G$.

\ \\
\textbf{Certificates.} Given any graph property ${\cal P}$ and a graph $G$, a \emph{certificate} of $G$ for ${\cal P}$ is a graph $G'$ on the same vertex set such that $G$ has ${\cal P}$ if and only if $G'$ has ${\cal P}$.

For any graph $G$ on vertex set $V$ and any property ${\cal P}$ a \emph{strong certificate} of $G$ for ${\cal P}$ is a graph $G'$ on vertex set $V$ such that for any graph $H$ on $V$, $G\cup H$ has ${\cal P}$ if and only if $G'\cup H$ has ${\cal P}$.

A certificate is said to be \emph{sparse} if the number of edges is ${\cal O}(n)$.

\ \\
\textbf{Semi-Streaming Algorithm.} A \emph{graph stream} of a graph $G$ is a sequence of the $m$ edges of $G$ in arbitrary oder.
A \emph{semi-streaming algorithm} $A$ gets a graph stream as an input and is restricted to use a space of at most ${\cal O}(n\cdot\mbox{polylog}\,n)$ bits. The algorithm may access the input stream for $P$ passes in a sequential one-way order. All algorithms considered in this paper use only $P=1$ pass. The per-edge processing time $T$ of $A$ we define to be the minimum time allowed between the revealing of two consecutive edges in the input stream. That definition of $T$ renders the definitions of previous papers more precisely, we give a discussion concerning that in Section \ref{PerEdgeTimeDiscussion}. There we also comment on the \emph{computing time} which denotes the total time required by $A$ to determine the property in question of the input graph.

\section{Discussion of Per-Edge Processing Time}\label{PerEdgeTimeDiscussion}
In previous papers about semi-streaming algorithms that consider the per-edge processing time $T$ (\cite{FeigenbaumKannanMcGregorSuriZhang04},\cite{FeigenbaumKannanMcGregorSuriZhang05},\cite{Zelke}), $T$ is used in an ambiguous way. While being used as the worst-case time to process a single edge on the one hand it is equally used on the other hand, even if not explicitly stated, as amortized time charged over the number of edges. In fact, if tools as dynamic trees or disjoint set data structures are utilized they give rise to amortized times since their time bounds are of amortized type, too. Processing the input edges is then assumed to be evenly spread over the whole computing time which is just $m\cdot T$.

This definition is not appropriate for a streaming algorithm: As Muthukrishnan\cite{Muthukrishnan} pointed out the computing time, i.e., the time to evaluate the property in question for items read in so far, is not the most important parameter of a streaming algorithm. What is more crucial is the maximum frequency of incoming items that can still be considered by the algorithm. That refers to the speed at which external storage devices can present their data content to a streaming algorithm and constitutes the frequency at which observed phenomena can be taken into account. To this aim it is desirable to maximize the possible rate of incoming items by postponing as much operations as possible to a point after which all items are received, possibly accepting a higher computing time.

To model this worthwhile property of a streaming algorithm $A$ we propose the definition of the per-edge processing time $T$ to be the minimum allowable time between two consecutive edges in the graph stream. The final determination of the property in question may require some postprocessing after reading all input edges. This time is considered in the computing time which incorporates the sum of the per-edge processing times of all edges and the postprocessing time.

\section{Computing Certificates and Buffering Edges}\label{MainSection}
To achieve our optimal per-edge processing times we exploit the general method of sparsification as presented by Eppstein et al.\cite{EppsteinGalilItalianoNissenzweig}. Feigenbaum et al.\cite{FeigenbaumKannanMcGregorSuriZhang05} pointed out how the results of \cite{EppsteinGalilItalianoNissenzweig} can be adopted for the semi-streaming model. Thus they received the formerly best bounds on $T$ for almost all problems considered in this paper. We refine their method to obtain an improvement of their results. For a comparison of our new bounds with the previous ones see Table \ref{Comparison}.

\begin{table*}[t]
\caption{\label{Comparison}Previously best per-edge processing times $T$ compared to our new  bounds}
\begin{tabular}{l@{\extracolsep{109pt}}cc}\hline
Problem & Previous Best $T$ & New $T$\\\hline
Connected components & ${\cal O}(\alpha(n))$ & ${\cal O}(1)$ \\[0.8ex]
Bipartition & ${\cal O}(\alpha(n))$ & ${\cal O}(1)$ \\[0.8ex]
\{2,3\}-vertex connectivity & ${\cal O}(\alpha(n))$ & ${\cal O}(1)$ \\
4-vertex connectivity & ${\cal O}(\log n)$ & ${\cal O}(1)$ \\
$k$-vertex connectivity & ${\cal O}(k^2n)$ & ${\cal O}(1)$ \\[0.8ex]
\{2,3\}-edge connectivity & ${\cal O}(\alpha(n))$ & ${\cal O}(1)$ \\
4-edge connectivity & ${\cal O}(n\alpha(n))$ & ${\cal O}(1)$ \\
$k$-edge connectivity & ${\cal O}(n\cdot\log n)$ & ${\cal O}(1)$ \\[0.8ex]
Minimum spanning forest & ${\cal O}(\log n)$ & $ {\cal O}(1)$ \\\hline
\end{tabular}\\[2ex]
All previous bounds are due to \cite{FeigenbaumKannanMcGregorSuriZhang05}, apart from $k$-vertex connectivity which is a result of \cite{Zelke}. $k$ is any constant, $\alpha(n)$ the inverse of Ackermann's function.
\end{table*}

\ \\
Due to the memory limitations of the semi-streaming model it is not possible to memorize a whole graph which is too dense, that is, if $m/n\gg\log n$. A way to determine graph properties without completely storing the graph is to find a sparse certificate $C$ of the graph for the property in question. Consisting of a linear number of edges the certificate can be stored within the memory restrictions and testing it answers the question for the original graph. The concept of certificates has been applied for the semi-streaming model in \cite{FeigenbaumKannanMcGregorSuriZhang05} and \cite{Zelke}. However, in \cite{Zelke} every input edge initiates an update of the certificate which is time-consuming and avoids a faster per-edge processing. 

To increase the manageable frequency of incoming edges, updating the certificate can be done not for every single edge but for a group of edges. While considering such a group of edges the next incoming edges can be buffered to compose the group for the following update. 

To permit this updating in groups of edges the utilized certificate must be a strong certificate, an assumption that is not required in \cite{Zelke}. That is because strong certificates obey two important attributes for any fixed graph property: Firstly, they behave transitively, that is, if $C$ is a strong certificate for $G$ and $C'$ is a strong certificate for $C$, then $C'$ is a strong certificate for $G$. Secondly, if $G'$ and $H'$ are strong certificates of $G$ and $H$ respectively, then $G'\cup H'$ is a strong certificate of $G\cup H$.

The technique of group-wise updating is used by Eppstein et al.\cite{EppsteinGalilItalianoNissenzweig} yielding fast dynamic algorithms and has been transferred to the semi-streaming model by Feigenbaum et al.\cite{FeigenbaumKannanMcGregorSuriZhang05}. 
The following theorem is a slightly extended version of their result augmented with space considerations. We will need details of the proof later on.

\begin{theorem}
\label{Theorem}
Let $G$ be a graph and let $C$ be a sparse and strong certificate of $G$ for a graph property ${\cal P}$. If $C$ can be computed in space ${\cal O}(m)$ and time $f(n,m)$, then there is a one-pass semi-streaming algorithm building $C$ of $G$ with per-edge processing time $T=f(n,{\cal O}(n))/n$.
\end{theorem}
\textbf{Proof.} We denote the edges of the input stream as $e_1,e_2,\ldots,e_m$ and the subgraph of $G$ containing the first $i$ edges in the stream as $G_i$. We inductively assume that we computed a sparse and strong certificate $C_{jn}$ of the graph $G_{jn}$ for $1\le j<\lfloor m/n\rfloor$ using a time of $f(n,{\cal O}(n))/n$ per already processed edge. During the computation of $C_{jn}$ we buffered the next $n$ edges $e_{jn+1},e_{jn+2},\ldots,e_{(j+1)n}$. 

Because of the properties of strong certificates  $T=C_{jn}\cup\{e_{jn+1},e_{jn+2},\ldots,e_{(j+1)n}\}$ is a strong certificate for $G_{(j+1)n}$. Since $C_{jn}$ is sparse, $T$ consists of ${\cal O}(n)$ edges as well. Computing $C_{(j+1)n}$ as a sparse and strong certificate of $T$ can be realized in a space linear in the space needed to memorize the edges of $T$, which is ${\cal O}(n\cdot\mbox{polylog}\,n)$ bits, without exceeding the memory limitation of the semi-streaming model. By transitivity $C_{(j+1)n}$ is a strong certificate of $G_{(j+1)n}$. A time of $f(n,{\cal O}(n))$ suffices to compute $C_{(j+1)n}$, hence the input edges can arrive with a time delay of $f(n,{\cal O}(n))/n$ building the group of the next $n$ edges to update the certificate after the computation of $C_{(j+1)n}$ is completed. 

Finally for $k=\lfloor m/n\rfloor$ the last group of edges $\{e_{kn+1},e_{kn+2},\ldots,e_m\}$ can simply be added to $C_{kn}$ to obtain a sparse and strong certificate of the input graph $G$ for the property ${\cal P}$. \hfill \qed

\ \\
To obtain our semi-streaming algorithms with optimal per-edge processing times, all that remains to do is to present the required certificates and to show in which time and space bounds they can be computed. At first glance it may seem surprising that Feigenbaum et al.\cite{FeigenbaumKannanMcGregorSuriZhang05} using the same technique of updating certificates with groups of edges do not meet the bounds we present in this paper. The reason is that they just observe that results of Eppstein et al.\cite{EppsteinGalilItalianoNissenzweig} can be transfered to the semi-streaming model. However, Eppstein et al. develop dynamic graph algorithms requiring powerful abilities: The algorithm must be able to answer a query for the subgraph of already read edges at any time and it must handle edge deletions. In the semi-streaming model the property is queried only at the end of the stream and there are no edge deletions. Thus we can drop both requirements for faster per-edge processing times.

\ \\
In the following the input graph for our semi-streaming algorithms is denoted by $G$ with $n$ vertices and $m$ edges as usual.
\subsection{Connected Components}
We use a spanning forest $F$ of $G$ as a certificate. $F$ is not only a strong certificate for connectivity it also has the same connected components as $G$. $F$ can be computed by a depth-first search in time and space of ${\cal O}(n+m)$ and is sparse by definition. Using Theorem \ref{Theorem} we get a semi-streaming algorithm computing a spanning forest of $G$ with per-edge processing time $T={\cal O}(1)$. To identify the connected components of $G$ in the postprocessing step we can run a depth-first search on the final certificate in time ${\cal O}(n)$. The resulting computing time is $m\cdot T + {\cal O}(n) = {\cal O}(n+m)$.

\subsection{Bipartition}
As a certificate for bipartiteness of $G$ we use $F^+$, which is a spanning forest of $G$ augmented with one more edge of $G$ inducing an odd cycle if there is any. If no such cycle exists $F^+$ is just a spanning forest. By \cite{EppsteinGalilItalianoNissenzweig} $F^+$ is a strong certificate of $G$ and sparse by definition. It can be computed by a depth-first search which is alternately coloring the visited vertices and is therefore able to find an odd cycle. To do so a time and space of ${\cal O}(n+m)$ suffices, yielding a semi-streaming algorithm with $T={\cal O}(1)$. On the final certificate we can run again a depth-first search coloring the vertices alternately in time ${\cal O}(n)$ during the postprocessing step. That produces a bipartition of the vertices or identifies an odd cycle in $G$ in a computing time of ${\cal O}(n+m)$. 

\subsection{$k$-Vertex Connectivity}\label{kVertexConn}
For $k$-vertex connectivity, $k$ being any constant, we use as a certificate of $G$ a subgraph $C_k$ which is derived by an algorithm presented by Nagamochi and Ibaraki\cite{NagamochiIbaraki}. $C_k$ can be computed in time and space of ${\cal O}(n+m)$, contains at most $kn$ edges and is therefore sparse. Beyond it, as a main result of \cite{NagamochiIbaraki} $C_k$ preserves the local vertex connectivity up to $k$ for any pair of nodes in $G$:
\begin{equation}\label{localVConn}
\kappa(x,y;C_k)\ge \min\{\kappa(x,y;G),k\} \hfill\forall x,y\in V\hfill
\end{equation}
This quality of $C_k$ leads to useful properties:

\begin{lemma}\label{sameSeps}
Every $l$-separator $S$ in $C_k$, $l<k$, is an $l$-separator in $G$ and its removal leaves the same connected components in both $C_k\setminus S$ and $G\setminus S$.
\end{lemma}
\textbf{Proof.} In $C_k\setminus S$ we find two nonempty, disjoint connected components $X$ and $Y$ with vertices $x\in X$ and $y\in Y$. Assume that $S$ is not an $l$-separator in $G$, therefore there exists a path $Z$ from $x$ to $y$ in $G\setminus S$. Let $x'$ be the last vertex on $Z$ in $X$ and $y'$ the first one in $Y$. The part of $Z$ between $x'$ and $y'$ we call $Z'$. In $C_k$ we find at most $l$ vertex-disjoint paths between $x'$ and $y'$, all of them using vertices of $S$. In $G$ these paths exist as well with the additional path $Z'$ which is vertex-disjoint from the other paths by construction. Therefore the local connectivity between $x'$ and $y'$ in $G$ exceeds that in $C_k$ contradicting property \ref{localVConn} of $C_k$.

Since $C_k\setminus S$ is a subgraph of $G\setminus S$ every connected component of $C_k\setminus S$ is included in one connected component of $G\setminus S$. Assume that $W$ is a connected component in $G\setminus S$ which contains two vertices $i$ and $j$ within different connected components of $C_k\setminus S$, namely $I\ni i$ and $J\ni j$. As in the first part of this proof we can find a path $Z$ from $i$ to $j$ in $W$ with $x'$ being the last vertex in $I$ and $y'$ the first one in $J$ on $Z$. We can deduce the same contradiction as above.\hfill \qed

\ \\So $C_k$ is usable for our purposes:
\begin{lemma}\label{StrongKConnCert}
$C_k$ is a strong certificate for $k$-vertex connectivity of $G$.
\end{lemma}
\textbf{Proof.} If $C_k\cup H$ is $k$-vertex connected then $G\cup H$ including $C_k\cup H$ as a subgraph is $k$-vertex connected as well. Assume for the proof of the converse direction that $G\cup H$ is $k$-vertex connected and $C_k\cup H$ is not. Then $C_k\cup H$ contains an $l$-separator $S$ for some $l<k$. After the removal of $S$ the remaining vertices of $C_k\cup H$ can be grouped into two nonempty sets $A$ and $B$, such that no edge joins a vertex of $A$ with a vertex of $B$. It is immediate that $H$ does not contain any edges between $A$ and $B$. 

Clearly, removing $S$ from $C_k$ produces the same sets $A$ and $B$, still with no edge joining them. The properties of $C_k$ shown in Lemma \ref{sameSeps} make sure that the removal of $S$ from $G$ leaves $A$ and $B$ without any joining edge, too. With $H$ having no edges between $A$ and $B$ the graph $G\cup H$ cannot be $k$-vertex connected. \hfill \qed

\ \\
Using Theorem \ref{Theorem} yields a semi-streaming algorithm computing a sparse and strong certificate of $k$-vertex connectivity in per-edge processing time $T={\cal O}(1)$. To test the final certificate for $k$-vertex connectivity in a postprocessing step we can use an algorithm of Gabow\cite{Gabow2000} on it. That algorithm runs in time ${\cal O}((k^{5/2}+n)kn)={\cal O}(kn^2)$ and, what is more important, uses a space linear in the number of edges of the final certificate, hence is respecting the memory constraints of the semi-streaming model. The resulting computing time is ${\cal O}(m+kn^2)$.

\subsection{$k$-Edge Connectivity}
We use the same $C_k$ as utilized in Section \ref{kVertexConn} produced by the algorithm of Nagamochi and Ibaraki presented in \cite{NagamochiIbaraki}, where it is shown that $C_k$ reflects the local edge-connectivity of $G$ in the following way:
\begin{equation}\label{localEConn}
\lambda(x,y;C_k)\ge \min\{\lambda(x,y;G),k\} \hfill\forall x,y\in V\hfill
\end{equation}
Therefore Lemma \ref{sameSeps} and Lemma \ref{StrongKConnCert} can be formulated and proven with respect to $l$-cuts, $l<k$, and $k$-edge connectivity. Accordingly we have a semi-streaming algorithm computing a strong and sparse certificate for $k$-edge connectivity using $T={\cal O}(1)$. To determine $k$-edge connectivity of the final certificate we can use an algorithm of Gabow\cite{Gabow1995} using a space linear in the number of edges of the final certificate. It takes a time of ${\cal O}(m+k^2n\log(n/k))$ which is also the resulting computing time of our semi-streaming algorithm.

\subsection{Minimum Spanning Forest}\label{MSF}
Let us first take a look at the algorithm we use as a subroutine for our semi-streaming algorithm computing an MSF of a given graph. We utilize the MST algorithm of Pettie and Ramachandran\cite{PettieRamachandran} which uses a space of ${\cal O}(m)$.    A remark on how we use an algorithm computing an MST to obtain an MSF we give below. The algorithm of \cite{PettieRamachandran} uses a time of ${\cal O}({\cal T}^*(m,n))$, where ${\cal T}^*(m,n)$ denotes the minimum number of edge-weight comparisons needed to find an MST of a graph with $n$ vertices and $m$ edges. The algorithm uses decision trees which are provably optimal but whose exact depth is unknown. Because of that the exact running time of the algorithm is not known even it is optimal. 

The currently tightest time bound for the MST problem is given by algorithms due to Chazelle\cite{Chazelle} and Pettie\cite{Pettie} that run in time ${\cal O}(m\cdot\alpha(m,n))$. Consequently the optimal algorithm of Pettie and Ramachandran\cite{PettieRamachandran} inherits this running time, ${\cal T}^*(m,n)={\cal O}(m\cdot\alpha(m,n))$. Based on the definition $\alpha(m,n)={\cal O}(1)$ if $m/n\ge \log n$. Therefore on a sufficiently dense graph the algorithm of \cite{PettieRamachandran} computes an MST in time ${\cal O}(m)$. 

Using this optimal algorithm as our subroutine we can find a semi-streaming algorithm with per-edge processing time $T={\cal O}(1)$ in the following way. We  use the technique described in Theorem \ref{Theorem} of merging a computed subgraph with buffered edges and then calculating a new subgraph of the merged graph while buffering the next group of edges. Unlike before we use groups of edges consisting of $r=n\cdot\log n$ edges instead of $n$. Such a number of edges can be memorized in the semi-streaming model using ${\cal O}(n\cdot\mbox{polylog}\,n)$ bits, even if weights are assigned to the edges which we assume to be storable in ${\cal O}(\mbox{polylog}\,n)$ bits each. 

By taking up the notation of the proof of Theorem \ref{Theorem}, $C_{jr}$ is the memorized MSF of the graph $G_{jr}$ made up of the edges $e_1,e_2,\ldots,e_{jr}$. We merge the buffered next $r$ edges with $C_{jr}$ to obtain $T=C_{jr}\cup\{e_{jr+1},e_{jr+2},\ldots,e_{(j+1)r}\}$. For the number $m_T$ of edges in $T$ we have $m_T\ge n\cdot\log n$ and therefore the optimal MST algorithm uses a time of ${\cal O}(m_T)$ to compute the MSF $C_{(j+1)r}$ of $T$. Since $m_T<2r$ the computation of $C_{(j+1)r}$ takes a time of ${\cal O}(r)$. To fill the buffer of the next $r$ edges in the meantime, the edges can arrive with a time delay of ${\cal O}(1)$.

It remains to show that what we compute in the described way is indeed an MSF of the input graph $G$. 
Every edge of $G_{jr}$ that is not in $C_{jr}$ is the heaviest on a cycle in $G_{jr}$ and cannot be in an MSF of $G_{jr}$. On the other hand $C_{jr}$ does not contain any dispensable edges since it includes no cycles: The removal of any edge from $C_{jr}$ produces two connected components in $C_{jr}$ whose vertices form a common connected component in $G_{jr}$. Therefore $C_{jr}$ forms an MSF of $G_{jr}$, inductively showing that we really obtain an MSF of $G$ in this manner.

Now we can state the computing time of our semi-streaming algorithm which depends on the density of the input graph $G$. If $G$ has at most $r=n\cdot\log n$ edges, all edges are read and buffered in time ${\cal O}(m)$ and then the optimal algorithm of Pettie and Ramachandran\cite{PettieRamachandran} computes an MSF in time ${\cal O}({\cal T}^*(m,n))$, producing a computing time of ${\cal O}({\cal T}^*(m,n))$, since $\Omega(m)$ is a lower bound for ${\cal T}^*(m,n)$.

If $G$ has more than $r$ edges we successively update an MSF with groups of edges. Note that, different from the described procedure in the proof of Theorem \ref{Theorem}, the last group of edges is not simply merged to the up to now computed $C_{\lfloor m/r \rfloor r}$. Instead the MSF of the merged graph is calculated to obtain the final MSF, which is also the MSF of the input graph, in the postprocessing step. We can fill the last group of edges up to a complete group of $r$ edges by using dummy edges weighted heavier than any edge in the input stream. This way we ensure that the last merged graph for the postprocessing with $m_f\ge r$ edges is sufficiently dense for the optimal MST algorithm running on it. So for the postprocessing time we have  ${\cal O}({\cal T}^*(m_f,n))={\cal O}(m_f\cdot\alpha(m_f,n))={\cal O}(m_f)$. Therefore the computing time is ${\cal O}(m)+{\cal O}(m_f)={\cal O}(m)$, which is trivially ${\cal O}({\cal T}^*(m,n))$.

\ \\
Let us give two minor remarks about the algorithm of Pettie and Ramachandran\cite{PettieRamachandran} we use. Firstly, the algorithm of \cite{PettieRamachandran} assumes the edge weights to be distinct. We do not require that property since ties can be broken while reading the input edges in a way described in \cite{EppsteinGalilItalianoNissenzweig}. Secondly, the algorithm of \cite{PettieRamachandran} works on connected graphs. Before running it, we can use a depth-first search to identify the connected components which are then processed separately. Identifying the connected components 
takes a time of ${\cal O}(m)={\cal O}({\cal T}^*(m,n))$, so the running time of our subroutine persists as well as the per-edge processing time of our semi-streaming algorithm.

\section{Discussion}\label{Discussion}
In this section we compare the obtained semi-streaming algorithms to algorithms determining the same properties in the classical RAM model allowing random access to all the edges of a graph without any memory constraints. 

First note that the presented semi-streaming algorithms have optimal per-edge processing times, that is, no semi-streaming algorithm exists allowing asymptotically shorter times: Every single edge must be considered to determine a solution for the problems considered in this paper, so a time of $\Omega(1)$ per edge is a lower bound for these problems.

Let us now take a look at the presented semi-streaming algorithms testing $k$-vertex and $k$-edge connectivity. For $k$-vertex connectivity with $k$ being a constant the fastest algorithm in the RAM model to date is due to Gabow\cite{Gabow2000} which runs in ${\cal O}(kn^2)$. Gabow obtains this result even in graphs with multiple edges by preprocessing the input graph with the algorithm of Nagamochi and Ibaraki\cite{NagamochiIbaraki} in time ${\cal O}(m)$ producing a running time of ${\cal O}(kn^2+m)$ on graphs and multigraphs. This asymptotically equals our computing time, which is not surprising since we use Gabow's algorithm as our subroutine. The same situation we find when looking at $k$-edge connectivity. Our achieved computing time of ${\cal O}(m+k^2n\log(n/k))$ is asymptotically as fast as the fastest algorithm in the RAM model due to Gabow\cite{Gabow1995} which we use as a subroutine. So both our connectivity algorithms have a computing time that is asymptotically the same as the fastest known corresponding algorithms in the RAM model.

It is possible that there are faster but still unknown algorithms in the RAM model for $k$-vertex and $k$-edge connectivity which cannot be utilized in the semi-streaming model because they consume to much space. The converse is true for the problems of finding connected components, a bipartition and an MSF of a given graph. The presented semi-streaming algorithms have asymptotically the same computing time as the fastest possible algorithms in the RAM-model. That can easily be seen for connected components and bipartition: We obtain in each case a computing time of ${\cal O}(n+m)$ which is trivially a lower bound for any algorithm in the RAM model solving these problems. For computing an MSF we get a computing time of ${\cal O}({\cal T}^*(m,n))$, where ${\cal T}^*(m,n)$ is the lower time bound for any RAM algorithm. 

For the asymptotic time needed to determine a solution there is no difference for $k$-edge and $k$-vertex connectivity between the currently fastest algorithms in the RAM model and the presented semi-streaming algorithms. Unless faster connectivity algorithms in the RAM model are developed there is no demand for a random access to the edges and for a memory exceeding ${\cal O}(n\cdot\mbox{polylog}\,n)$ bits. For computing the connected components, a bipartition and an MSF such a demand will never emerge since the presented semi-streaming algorithms have optimal computing times. The RAM model cannot capitalize on its mighty potential of unlimited memory and random access to beat the computing times of the weaker semi-streaming model.

\ \\
We close this section by indicating a tradeoff between memory and time when computing an MSF in the semi-streaming model. If the memory constraint of the semi-streaming algorithm is reduced from ${\cal O}(n\cdot\mbox{polylog}\,n)$ to ${\cal O}(n\cdot\log^{2-\varepsilon} n)$ bits, only $s=o(n\cdot\log n)$ edges can be memorized. So the optimal MST algorithm we use as a subroutine needs a time of ${\cal O}({\cal T}^*(s,n))$. Provided that ${\cal T}^*(s,n)=\omega(s)$ we obtain a per-edge processing time of $\omega(1)$ and therefore a computing time of $\omega(m)$. Both bounds are significantly larger than the corresponding ones when ${\cal O}(n\cdot\mbox{polylog}\,n)$ bits of memory are permitted. However, if it turns out that ${\cal T}^*(m,n)={\cal O}(m)$ for any $m$, it suffices to store $\Theta(n)$ edges to obtain both optimal per-edge and computing time in the semi-streaming model.

\section{Conclusion}\label{Conclusion}
We presented semi-streaming algorithms for computing the connected components, a bipartition, the $k$-vertex and $k$-edge connectivity for any constant $k$ and an MSF of a given graph. The presented per-edge processing times $T$ surpass former semi-streaming algorithms and are optimal because they are constant. All introduced semi-streaming algorithms are asymptotically as fast as the fastest corresponding algorithms in the RAM model. For connected components, bipartition and MSF we actually achieve the time bounds of the best possible RAM algorithms.

The main idea for our semi-streaming algorithms is quite simple: A sparse memorized subgraph is merged with buffered edges and while computing a sparse subgraph of the merged one the next edges are buffered. We believe this idea to be fruitful for other graph problems as well when tackling them without random access and within the memory constraints of the semi-streaming model.

\end{document}